\documentclass[twocolumn]{jpsj2} 
\title{Critical Nature of Non-Fermi Liquid in Spin $3/2$ Multipolar Kondo Model}

\author{Kazumasa \textsc{HATTORI}}

\inst{Division of Materials Physics, Department of Materials Engineering Science, Graduate School of Engineering Science, Osaka University, Toyonaka, Osaka 560-8531, Japan}

\abst{A multipolar Kondo model of an impurity spin $S_I=3/2$ interacting with conduction electrons with spin $s_c=3/2$ is investigated using boundary conformal field theory. A two-channel Kondo (2CK) -like non-Fermi liquid (NFL) under the particle-hole symmetry is derived explicitly using a ``superspin absorption'' in the sector of a hidden symmetry, SO(5). We discuss the difference between the usual spin-$1/2$ 2CK NFL fixed point and the present one. In particular, we find that, unlike the usual 2CK model, the low temperature impurity specific heat is proportional to temperature.}

\kword{non-Fermi liquid, boundary conformal field theory, Kondo effect, SO(5), numerical renormalization group}

\begin{document}
\maketitle
 An intuitive understanding of a prototype of non-Fermi liquid (NFL), a multichannel Kondo model, was proposed by Nozi\`eres and Blandin about 25 years ago.\cite{NB} The understanding of the multichannel Kondo model was obtained by various methods, such as Bethe-ansatz\cite{Wiegman, Andrei}, boundary conformal field theory (BCFT)\cite{AffLud1,AffLud2,AffLud3,AffLud4,2IK}, numerical renormalization group (NRG)\cite{Cragg,Pang}, and so on. In particular, we can compare the NRG finite size spectra and those of the BCFT even if we have NFL spectra and the agreements are excellent.\cite{Kim2,AffLud4,2IK} 

 In the realistic model of diluted f-electron systems, a candidate for a two-channel Kondo (2CK) model is the quadrupolar Kondo model\cite{Cox} for U and Ce based alloys, in which non-Kramers doublet in $f^2$ configuration plays an important role. Several models that have a NFL property were proposed, in which the $\Gamma_8$ conduction electrons interact with a localized $f^1$ or $f^2$ crystalline-electric-field state under cubic symmetry\cite{Koga1,KusuKura,Koga2,Kim1}. These models may be relevant for dilute alloys of Ce$^{3+}$ or U$^{4+}$ ions, such as La$_{1-x}$Ce$_x$Cu$_2$Si$_2$\cite{Andraka}, UCu$_{5-x}$Pd$_x$\cite{Andrade} among others. In general, NFL behaviors are observed in real materials as transient phenomena, which are controlled by temperature, pressure, alloying and so on.

 In this Letter, we reexamine an impurity Kondo model in which a localized f-electron with $\Gamma_8$ symmetry interacts with $\Gamma_8$ conduction electrons under cubic symmetry. This model was proposed as that of Ce$_x$La$_{1-x}$B$_6$\cite{Koga1}. The earlier NRG calculations confirmed the existence of a NFL fixed point that is unstable  against the particle-hole (PH) symmetry breaking. Indeed, Ce$_x$La$_{1-x}$B$_6$ is not located near this NFL fixed point, so that the property of this NFL may not be relevant to the case of Ce$_x$La$_{1-x}$B$_6$. However, the NFL would be realized in the system with small quadrupolar interactions that break PH symmetry. Theoretically, the origin of the NFL was thought to be mysterious from the conventional BCFT point of view. The NRG energy spectrum was similar to that of 2CK model but the origin was unknown.

Although it is difficult to control PH asymmetry in experiments, for both the pursuit of new materials and theoretical interest, it is worthwhile to study the detailed properties of this NFL by BCFT. It is noted that a BCFT approach is a powerful tool for answering {\it why} NFL behaviors emerge in various impurity problems. The same situation occurs in the case of the two-impurity Kondo model (the NFL is unstable against PH asymmetry of conduction electrons), which has been extensively investigated so far\cite{Jones}. Roughly speaking, the quantum critical phenomena near the antiferromagnetic critical point at zero temperature\cite{Stewart} can be seen as a kind of a lattice generalization of the two-impurity Kondo model (however, it is not so simple). The PH asymmetry in a generalized 2CK model 
was also discussed\cite{Pan2}. In this case, in the presence of a double tensor interaction that breaks the PH symmetry, the system flows to a Fermi liquid.

 We assume the following points for the local f-electron state: (i) cubic symmetry, (ii) $f^1$ $\Gamma_8$ ground state configuration, (iii) the lowest excited state is f$^0$ and/or $f^2$ $\Gamma_1$ singlet configuration. 
 Under these assumptions, we can map the $\Gamma_8$ index to a pseudospin-$3/2$ representation as $|\Gamma_{8,{\pm \frac{3}{2}}}\rangle=\mp(| \pm \frac{3}{2}\rangle+\sqrt{5}|\mp \frac{5}{2}\rangle)/\sqrt{6}$ and $|\Gamma_{8,{\pm \frac{1}{2}}}\rangle=\pm| \pm \frac{1}{2}\rangle$, where $|j_z\rangle$ is a state with the total angular momentum $J=5/2$ and its z-component $j_z$.

We start with the following pseudospin-$3/2$ Kondo model (see the derivation in ref. 13) in 1-dimension considering the s-wave scattering at the impurity site\cite{AffLud2}:
\begin{eqnarray}
H&=&H_0+\sum_{m=\rm dip,quad,oct}H_{m},\label{H}\\
H_0&=&\frac{iv_F}{2\pi}\sum_{\mu=\pm \frac{3}{2},\pm\frac{1}{2}}\int dx\psi_{\mu}^{\dagger}(x)\partial \psi_{\mu}(x),\\
H_{m}&=& J_{m}\sum_{\mu,\nu=\pm \frac{3}{2},\pm\frac{1}{2}}\psi_{\mu}^{\dagger}(0)({\bf x}_c^m)_{\mu\nu}\psi_{\nu}(0)\cdot {\bf X}_I^m,
\end{eqnarray}
where we introduce left-moving fermion annihilation operators $\psi_{\mu}(x)$, Fermi velocity $v_F$ and spin $3/2$ dipolar (${\bf x}_c^{\rm dip}={\bf s}_c$), quadrupolar (${\bf x}_c^{\rm quad}={\bf q}_c$) and octupolar (${\bf x}_c^{\rm oct}={\bf t}_c$) matrices of the conduction electron (similar definitions for ${\bf X}_I^m$ of the impurity).\cite{Relation} $J_{m}$ is the coupling constant of each multipoles. It is noted that under the assumptions (i)-(iii) the interactions are isotropic in the pseudospin space.

The ``conformal embedding'' often found in literatures is SU(4)$\to$ SU(2)$_{10}\otimes$ SU(2)$_4$.\cite{Kim2}
 The SU(2)$_{10}$ corresponds to a spin current $ {{ \mbox{\boldmath $ \mathcal J$}}}(x)$\cite{J}, i.e., {\it dipole}. The other SU(2) current, SU(2)$_4$, is an axial charge (AC) current $\mbox{\boldmath $\mathcal Q$}(x)$\cite{Q}. In this embedding, $H_0$ can be written in the following Sugawara form:
\begin{eqnarray}
\frac{l}{\pi v_F}H_0=\sum_{n=-\infty}^{\infty}\Big(\frac{1}{4}:\mbox{\boldmath $\mathcal Q$}_{-n}\cdot\mbox{\boldmath $\mathcal Q$}_{n}: + \frac{1}{12}:{\mbox{\boldmath $\mathcal J$}}_{-n}\cdot 
{\mbox{\boldmath $\mathcal J$}}_{n}:\Big), \label{Hboson}
\end{eqnarray}
where 
${\mbox{\boldmath $\mathcal J$}}_{n}\ {\rm and}\ {\mbox{\boldmath $\mathcal Q$}}_{n}$ are the Fourier components of ${\mbox{\boldmath $\mathcal J$}}(x)\ {\rm and}\ {\mbox{\boldmath $\mathcal Q$}}(x)$, respectively. We set the system size to $2l$. $:A:$ indicates the normal ordering of the operator $A$. The energy eigenvalues of the right hand side of eq. (\ref{Hboson}) for primary states that form conformal towers\cite{Itz} are given by
\begin{eqnarray}
E(q,j)=\frac{q(q+1)}{4}+\frac{j(j+1)}{12},\label{E0}
\end{eqnarray}
 where $(q=0,1/2, 1)$ and $(j=0,1/2, \cdots, 5)$. The energies of the descendant states\cite{Itz} are given by $E(q,j)+n$, where $n$ is a positive integer and indicates the PH excitations from the ground state.\cite{AffLud3} The descendant states generally have different quantum numbers from those of the corresponding primary state in the same conformal tower.
 The form of eq. (\ref{Hboson}) is suitable to the case of $J_{\rm quad}=J_{\rm oct}=0$. In this case, the impurity spin can be absorbed into ${\mbox{\boldmath $\mathcal J$}}_{n}$\cite{AffLud1} and the predicted NFL spectra are in complete agreement with the NRG results\cite{Kim2,Koga1,Koga2}.

In the presence of quadru- and octupolar interactions, the above SU(2)$_{10}$ absorption is not applicable.
As noted by Wu {\it et al.} the spin $3/2$ fermionic system has an exact SO(5) symmetry under some conditions\cite{Wu1}. In the following, we transform the Hamiltonian (\ref{H}) into the SO(5) language, i.e., the embedding is SU(4)$\to $SO(5)$_2\otimes$ SU(2)$_4$.

First, quadrupolar matrices of $s_c=3/2$ have the following forms:
\begin{eqnarray}
q_c^{3z^2-r^2}&\equiv&\frac{1}{\sqrt{3}}(3s_c^zs_c^z-\frac{3}{2}\frac{5}{2}\hat{\bf 1})\equiv \sqrt{3}\Gamma^4 \label{qz},\\
q_c^{x^2-y^2}&\equiv&\frac{1}{\sqrt{2}}(s_c^xs_c^x-s_c^ys_c^y)\equiv-\sqrt{3}\Gamma^5,\\
q_c^{xy}&\equiv&s_c^xs_c^y+s_c^ys_c^x\equiv -\sqrt{3}\Gamma^1,\\
q_c^{yz}&\equiv&s_c^ys_c^z+s_c^zs_c^y\equiv\sqrt{3}\Gamma^3,\\
q_c^{zx}&\equiv&s_c^zs_c^x+s_c^xs_c^z\equiv\sqrt{3}\Gamma^2,\label{qzx}
\end{eqnarray}
where $\hat{\bf 1}$ is an identity matrix and the spin $3/2$ operators $s_c^i\ (i=x,y,z)$ are constructed on the basis of $^{\dagger}(\psi_{\frac{3}{2}},\psi_{\frac{1}{2}},\psi_{\frac{-1}{2}},\psi_{\frac{-3}{2}})$. We have introduced five Dirac matrices $\Gamma^a\ (1\le a \le 5)$.
From eqs. (\ref{qz})-(\ref{qzx}), we can define ten SO(5) generators $\Gamma^{ab}$ as $\Gamma^{ab}\equiv \frac{1}{2i}[\Gamma^a, \Gamma^b]$. These generators satisfy the SO(5) commutation relations:
\begin{eqnarray}
[\Gamma^{ab}, \Gamma^{cd}]=-2i(\delta_{bc}\Gamma^{ad}-\delta_{ac}\Gamma^{bd}-\delta_{bd}\Gamma^{ac}+\delta_{ad}\Gamma^{bc}).
\end{eqnarray}

The important point is that $H_{\rm dip}$ and $H_{\rm oct}$ are written by the ten $\Gamma^{ab}$ when the condition $J_{\rm dip}=J_{\rm oct}$ is satisfied. Thus, under this condition, each multipolar Hamiltonian can be written in the SO(5)'s vector $n^a_I$ ($n_c^a$) and generators $L^{ab}_I$ ($L_c^{ab}$) of the impurity (conduction electron), as follows:
\begin{eqnarray}
H_{\rm quad}&=& 12J_{\rm quad}\sum_{a=1}^5n_c^a(0)n^a_I,\label{Hq}\\
H_{\rm dip}+H_{\rm oct}&=& 5J_{\rm dip}\sum_{a=1}^5\sum_{b>a}^5L_c^{ab}(0)L^{ab}_I,\label{Hdo}
\end{eqnarray}
 where 
\begin{eqnarray}
n_c^a(x)&\equiv& \frac{1}{2}\sum_{\mu\,\nu=\pm \frac{3}{2},\pm\frac{1}{2}}\psi_{\mu}^{\dagger}(x)(\Gamma^a)_{\mu\nu}\psi_{\nu}(x),\label{nc}\\
L_c^{ab}(x)&\equiv&-\frac{1}{2}\sum_{\mu,\nu=\pm \frac{3}{2},\pm\frac{1}{2}}\psi^{\dagger}_{\mu}(x)(\Gamma^{ab})_{\mu\nu}\psi_{\nu}(x).\label{Lc}
\end{eqnarray}
\begin{table}[t]
	\begin{tabular}{cccc}
         \hline
         $q$ & $j$ & {\bf u} & $E(q,{\bf u}) \ (E(q,j))$\\
         \hline
         \hline
         $0$ & $0$ & $\bf 1$ & $0$\\
         \hline
         $1/2$ & $3/2$ & $\bf 4$ & $1/2$\\
         \hline
         $1$ & $2$ & $\bf 5$ & $1$\\
         \hline
         $0$ & $3$ & $\bf 10$ & $1$\\
         $0$ & $1$ &      & $1$\\
         \hline
         $1$ & $0$ & $\bf 1$  & $1$\\
         \hline
         $3/2$ & $3/2$ & $\bf 4$  & $3/2$\\
         $1/2$ & $3/2$ & $\bf 4$  & $3/2$\\
         \hline
         $1/2$ & $7/2$ & $\bf 16$ & $3/2$\\
         $1/2$ & $5/2$ &      & $3/2$\\
         $1/2$ & $1/2$ &      & $3/2$\\
         \hline
        \end{tabular}
\caption{Low-energy spectrum of free Hamiltonian for nondegenerate ground state. The first and the third columns are the AC and SO(5) indices, respectively. The second column is the eigenvalue of the spin current in eq. (\ref{E0}). $E(q,{\bf u})$ and $E(q,j)$ are measured from the ground state using eqs. (\ref{E1}) and (\ref{E0}), respectively.}
\label{tbl-1}
\end{table}

\begin{figure}[t]
  \begin{center}
    \includegraphics[width=.25\textwidth]{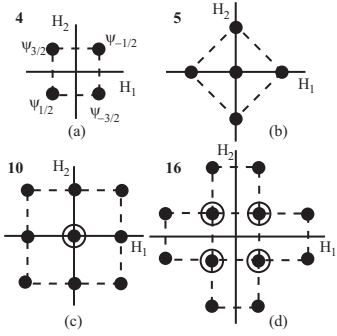}
  \end{center}
\caption{$H_1-H_2$ diagrams associated with the SO(5) multiplies that enter the low-energy spectrum, where $2H_1\equiv \Gamma^{15}$ and $2H_2\equiv \Gamma^{23}$ forming the Cartan subalgebra of SO(5) group. In (a), the corresponding components of the spinor are indicated.}
\label{fig-1}
\end{figure}

The free Hamiltonian $H_0$ in (\ref{H}) is expressed using the SO(5) generators
\begin{eqnarray}
\frac{l}{\pi v_F}H_0=\sum_{n=-\infty}^{\infty}\Big(\frac{1}{4}:\mbox{\boldmath $\mathcal Q$}_{-n}\cdot\mbox{\boldmath $\mathcal Q$}_{n}: + \frac{1}{8}:{\mbox{\boldmath $\mathcal L$}}_{-n}\cdot 
{\mbox{\boldmath $\mathcal L$}}_{n}:\Big), \label{Hboson2}
\end{eqnarray}
where $({\mbox{\boldmath $\mathcal L$}}_{n})^{ab}$ is the n-th Fourier component of SO(5) current $L_c^{ab}(x)$: ${\mbox{\boldmath $\mathcal L$}_n}=({\mathcal L^{12}_{n}}, {\mathcal L^{13}_{n}},\cdots ,{\mathcal L^{35}_{n}}, {\mathcal L^{45}_{n}})$ and satisfies a level 2 SO(5) Kac-Moody algebra. The energy eigenvalues $E(q,{\bf u})$ for the primary states are expressed by the Casimir operators in each sector
\begin{eqnarray}
E(q,{\bf u})=\frac{q(q+1)}{4}+\frac{\bar{c}_{\bf u}}{8},\label{E1}
\end{eqnarray}
where $q=0,1/2, 1$ for the AC sector (this is as same as in eq. (\ref{E0})), and $\bf u=1\ {\rm(identity)},\ 4\ {\rm (spinor)}\ ,5\ {\rm (vector)}$ for the SO(5) sector. ${\bf u}$ indicates the dimension of the representation in the SO(5), see Fig. \ref{fig-1}. There are three primary fields in both the AC and the SO(5) sectors.
The values of $\bar{c}_{\bf u}$ are $\bar{c}_{\bf 1}=0,\ \bar{c}_{\bf 4}=5/2\ {\rm and }\  \bar{c}_{\bf 5}=4$. The noninteracting energy spectra for the nondegenerate ground state are shown in Table \ref{tbl-1} together with the indices of $j$ in eq. (\ref{E0}). We can see that both eqs. (\ref{E0}) and (\ref{E1}) can reproduce the spectrum of free $s_c=3/2$ fermions.

Next, we consider eqs. (\ref{Hq}), (\ref{Hdo}) and (\ref{Hboson2}) together. It is noted that the total Hamiltonian (\ref{H}) is written as

\begin{eqnarray}
\frac{l}{\pi v_F}H&=&\sum_{n=-\infty}^{\infty}\Big(\frac{1}{4}:\mbox{\boldmath $\mathcal Q$}_{-n}\cdot\mbox{\boldmath $\mathcal Q$}_{n}: + \frac{1}{8}:{\mbox{\boldmath $\mathcal L$}}_{-n}^{'}\cdot 
{\mbox{\boldmath $\mathcal L$}}_{n}^{'}:\Big)\nonumber\\
&&+\frac{12lJ_{\rm quad}}{\pi v_F}\sum_{a=1}^5n_c^a(0)n^a_I+\rm const.,\label{HH}
\end{eqnarray}
where
\begin{eqnarray}
{\mathcal L'}_{n}^{ab}\equiv{\mathcal L}_{n}^{ab}+\frac{20lJ_{{\rm dip}}}{\pi v_F}L_I^{ab}.
\end{eqnarray}
The values at possible fixed points, the $J_{\rm dip}^*$, $J^*_{\rm quad}$ and $J^*_{\rm oct}$ are determined (assumed) as
\begin{eqnarray}
   J_{\rm dip}^*=J_{\rm oct}^*=\frac{\pi v_F}{20l}{\rm,\ \ \ \ and}\ \ \ J^*_{\rm quad}=0.\label{Jc}
\end{eqnarray}
 The case in which $J_{\rm quad}\not=0$ is discussed later. At these values of the couplings, the impurity SO(5) ``superspin'' can be absorbed into the conduction electron ``superspin'' current, generating the new SO(5) ``superspin'' $ {\mbox{\boldmath $\mathcal L$}}_{n}^{'}$. This impurity absorption does not affect the SO(5) Kac-Moody algebra, except the gluing conditions in Table \ref{tbl-1}, which is much the same as in the case of the multichannel Kondo model\cite{AffLud1}.

 At this stage, there is a need to introduce a suitable fusion rule to generate the nontrivial spectrum. Because the interaction is only in the SO(5) sector, any fusion in the AC sector is unphysical. An important point is that the impurity is described in the spin $3/2$ representation (i.e., dimension 4). We find that the desired fusion is a direct product of $\bf 4$ representation in the SO(5) sector to each state in the spectra of the free Hamiltonian. To execute this, the following formulae are useful (easily deducible from Fig. \ref{fig-1}): $
\bf 1\otimes 4= \bf 4,\ 
\bf 4\otimes 4= \bf 1\oplus 5\oplus 10\ {\rm and}\ 
\bf 4\otimes 5= \bf 4\oplus 16$.

Thus, even after absorbing the impurity ``superspin'', we can calculate the energy spectra at this fixed point by using eq. (\ref{E1})
. 
The obtained spectra are shown in Table \ref{tbl-2}(a). As shown by previous NRG studies\cite{Koga1, KusuKura}, the NFL spectra are the same as those of the 2CK NFL if the AC sector in the present model and the spin sector in the 2CK model are interchanged. Indeed, the existence of a SO(5) symmetry in the 2CK model has been pointed out before in literatures.\cite{AffLud4}

\begin{table}[t]
\begin{center}
	\begin{tabular}{cccccccc}
         \hline
         (a)&$q$  & {\bf u} & $E(q,{\bf u})$  & (b) &  $q$ & {\bf u} & $\Delta$\\
         \hline
         \hline
         &$1/2$ & $\bf 1$ & $0$       &     &  $0$ & $\bf 1$ & $0$     \\
         &$0$ &   $\bf 4$ & $1/8$     &     &  $0$ & $\bf 5$ & $1/2$   \\
         &$1/2$ & $\bf 5$ & $1/2$     &     &  $1/2$ & $\bf 4$ & $1/2$ \\
         &$1$ & $\bf 4$   & $5/8$     &     &  $1/2$ & $\bf 4$ & $1/2$ \\
         &$3/2$ & $\bf 1$ & $1$       &     &  $1$   & $\bf 1$ & $1/2$ \\
         &$1/2$ & $\bf 10$ & $1$      &     &  $1$ & $\bf 5$  & $1$    \\
         &$1/2$ & $\bf 1$ & $1$       &     &  $0$ & $\bf 10$  & $1$   \\
         \hline
        \end{tabular}
\caption{(a) Low energy spectrum of NFL fixed point with $E(q,{\bf u})\le 1$. In the third column, the ground state energy $3/16$ is subtracted. (b) Operator contents at the NFL fixed point. We show the operators with $\Delta \le 1$ where $\Delta$ means the scaling dimension of the corresponding operator. 
}
\label{tbl-2}
\end{center}
\end{table}

%
%

The operator contents at the NFL fixed point are obtained by applying a double fusion\cite{AffLud1} in the SO(5) sector (see Table \ref{tbl-2}(b)). Again, we obtain the same scaling dimensions $\Delta$ as those at the 2CK fixed point with the above-mentioned interchange. The operator with $(q, {\bf u})=(0,{\bf 5})$ is the SO(5) vector $\mbox{\boldmath $\mathcal \phi$}_{\rm SO(5)}$, which corresponds to the quadrupolar operator in the original spin $3/2$ representation. This operator has $\Delta=1/2$, so that the corresponding local quadrupolar susceptibility diverges logarithmically at low temperatures. This is consistent with the result of NRG studies \cite{Koga1, KusuKura}. The local charge and pair field susceptibility would diverge at low temperatures, because the operator $(q, {\bf u})=(1,{\bf 1})$, $\mbox{\boldmath $\mathcal \phi$}_{\rm AC}$, which is the AC vector, has $\Delta=1/2$. The dipolar susceptibility in addition to the octupolar one, is classified in $\bf 10$ representation with $\Delta=1$. This means that the dipolar and the octupolar susceptibilities $\chi$ are not singular, which is also consistent with the NRG calculation\cite{Koga1,KusuKura}.

Next we discuss the stability of the NFL fixed point against various perturbations.

a) In the presence of a uniaxial distortion, the term $h_Qn_I^4$, which breaks the SO(5) symmetry, appears in the effective Hamiltonian. This term allows $\mbox{\boldmath $\mathcal \phi$}_{\rm SO(5)}$ with $\Delta=1/2$ to appear. Thus, the conjugate field $h_Q$ becomes a relevant perturbation and the NFL fixed point becomes unstable.

b) In the presence of a potential scattering at the impurity site $V\sum_{\mu}\psi_{\mu}^{\dagger}(0)\psi_{\mu}(0)=VQ_z(0)+\ \rm const.$, the SU(2) symmetry in the AC sector is broken. In this case, $\phi_{\rm AC}^3$, the component $q_z=0$ of $\mbox{\boldmath $\mathcal \phi$}_{\rm AC}$ with $\Delta=1/2$, can appear in the effective Hamiltonian. Again, the NFL fixed point becomes unstable. The PH symmetry can also be broken by the quadrupolar interaction of eq. (\ref{Hq}). Indeed, the system flows into the Fermi liquid fixed point of SU(4) Coqblin-Schrieffer model without PH symmetry\cite{Koga1}. 

c) The exchange anisotropy in the SO(5) sector is irrelevant, because the marginal operator $(q,{\bf u})=(0,{\bf 10})$ cannot appear under the time-reversal symmetry as discussed in Ref. 8 (that is, our assumption $J_{\rm dip}=J_{{\rm oct}}$, does not affect the present result).\cite{COMMENT}

\begin{figure}[t!]
  \begin{center}
    \includegraphics[width=.4\textwidth]{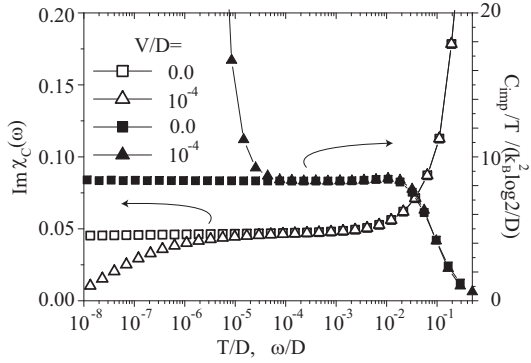}
  \end{center}
\caption{$C_{\rm imp}/T$ vs $T$ and ${\rm Im}\chi_c(\omega)$ vs $\omega$. $V$ is the strength of potential scattering at the impurity site. The parameters used are $J_{\rm dip}=J_{\rm oct}=0.1D$ and $J_{\rm quad}=0.0D$, where $D$ is the half of the bandwidth of conduction electrons.}
\label{fig-2}
\end{figure}

In early NRG studies\cite{Koga1,KusuKura}, the NFL fixed point was considered to be similar to the 2CK fixed point. But the question arises, `` Is the present NFL really equivalent to the 2CK ?'' To verify this, we consider the effective Hamiltonian near the NFL fixed point. If the present NFL and the NFL of the 2CK were equivalent, the leading irrelevant operator would be $\mbox{\boldmath $\mathcal Q$}_{-1}\cdot\mbox{\boldmath $\mathcal \phi$}_{\rm AC}$ with $\Delta=1/2+1=3/2$. This operator, however, is not physically adequate because the impurity absorption occurs in the SO(5) sector. The low-energy effective Hamiltonian should be made by using operators in the SO(5) sector and these should transform as a singlet\cite{AffLud1}. 

Because ${\mathcal L}^{ab}_{-1}\phi^c_{\rm SO(5)}\ (1\le a,b,c\le 5)$ type operators with $\Delta=3/2$ cannot become a singlet, the leading irrelevant operator should be the energy-momentum operator $\sum_{ab}:L_c^{ab}(0)L_c^{ab}(0):$ with $\Delta=2$. This leads us to an important conclusion: {\it impurity specific heat does not diverge at low temperatures} unlike in the 2CK case, {\it but shows a linear temperature behavior}. As we mentioned above, $\chi$ is not singular, so the Wilson ratio $R_W=(\delta\chi/\chi)/(\delta C/C)$ is calculated using the conformal charge in each of sectors $c_{\rm SO(5)}$ and $c_{\rm AC}$ as $R_W=(c_{\rm SO(5)}+c_{\rm AC})/c_{\rm SO(5)}=(5/2+3/2)/(5/2)=8/5$($C$: specific heat).\cite{AffLud3} This result is different from that of the 2CK case: $R_W=8/3$.

Finally, to check the new results above, we show the results of the NRG calculations of the impurity specific heat $C_{\rm imp}$ and the z-component of the dynamical AC susceptibility for localized conduction electrons $\chi_c(\omega)$ at zero temperature in Fig. {\ref{fig-2}}. Note that $\chi_c(\omega)$ is the dynamical charge susceptibility of the conduction electrons at the impurity site. We used logarithmic discretization parameter $\Lambda=3$ and kept 400 states at each NRG step. It was confirmed that $C_{\rm imp}/T$ is constant at low temperatures and that Im$\chi(0)\not=0$ indicates the divergence of the conduction electron's charge susceptibility. We can see that the divergence of the charge susceptibility is suppressed when PH symmetry is broken ($V\not=0$ case). We also obtained the residual impurity entropy $S_0=k_B\log \sqrt{2}$, which is the same value as that at the 2CK NFL fixed point. An increase in the data $V/D=10^{-4}$ of the $C_{\rm imp}/T$ is the natural consequence of the entropy release from $k_B\log \sqrt{2}$ to $0$.

In summary, we have investigated a multipolar Kondo model with $S_I=3/2$ and $s_c=3/2$ using BCFT and NRG. The 2CK-like NFL fixed point observed in the earlier NRG calculation is explicitly derived using the ``superspin'' absorption in a hidden symmetry SO(5). We find that the leading irrelevant operator at the NFL fixed point is ``Fermi liquid-like'' in contrast to its 2CK-like NFL spectra. All predictions are consistent with earlier works and the present NRG calculation. In particular, the low temperature impurity specific heat is proportional to temperature, the Wilson ratio $R_W=8/5$ and the local charge susceptibility of conduction electrons diverges at zero temperature. These results remarkably distinguish the present NFL from the 2CK model.

\vspace{.5cm}
The author would like to thank A. Yotsuyanagi, T. Takimoto, H. Kohno and K. Miyake for valuable comments. This work was supported by the 21st Century COE Program (G18) of the Japan Society for the Promotion of Science.
The author is supported by the Research Fellowships of JSPS for Young Scientists.


\begin{thebibliography}{99}
\bibitem{Q} The axial charge density operator is defined as \\
${Q}_z(x) = \sum_{\mu} (\psi^{\dagger}_{\mu}(x)\psi_{\mu}(x)-1/2)$,\\
${Q}_+(x) = (Q_-(x))^{\dagger}=\psi^{\dagger}_{3/2}(x)\psi^{\dagger}_{-3/2}(x)-\psi^{\dagger}_{1/2}(x)\psi^{\dagger}_{-1/2}(x)$.
\bibitem{COMMENT} In the $J_{\rm oct}=0$ case, the impurity spin seems to be absorbed as if $S_I=1/2$ with an even-odd interchange of the NRG iteration number, resulting SU(2) NFL fixed point\cite{Koga1}. Although the precise fusion is not clear, if we admit $S_I=1/2$ fusion, $J_{\rm oct}$ might be irrelevant at the fixed point. This might suggest the existence of 1st order line between two NFL fixed points.
\end{thebibliography}
\end{document}